\documentstyle[11pt]{article}
\newcommand{\blankline}{\vskip .3cm}
\newcommand{\f}{\begin{equation}}
\newcommand{\ff}{\end{equation}}

\begin{document}
\rightline{CGPG-96/94}
\blankline
\blankline
\blankline
\centerline{\LARGE The classical limit and the form of the 
hamiltonian constraint}
\blankline
\centerline{\LARGE in non-perturbative quantum general relativity}
\blankline
\blankline
\rm
\centerline{Lee Smolin${}^*$}
\blankline
 \centerline{\it  Center for Gravitational Physics and Geometry}
\centerline{\it Department of Physics}
 \centerline {\it The Pennsylvania State University}
\centerline{\it University Park, PA, USA 16802}
 \vfill
\centerline{September 4, 1996}
\vfill
\centerline{ABSTRACT}
It is argued that some approaches to non-perturbative  quantum 
general relativity lack a sensible continuum limit that reproduces
general relativity.  This may be true in spite of their being mathematically 
well defined diffeomorphism invariant quantum field theories that 
result from applying canonical quantization to general relativity.  The
basic problem is that generic physical states lack long ranged correlations, 
because the form of the state allows a division into spatial regions, 
such that no change in the physical  state in one region can be 
measured by observables restricted to another.  These disconnected 
regions have generically finite expectation value of physical volume, which 
means  that the theory has no long ranged correlations or massless particles.
One consequence of this is that the $ADM$   energy is unbounded 
from below, at least when that is defined with respect to a natural 
notion of quantum  asymptotic flatness and a   corresponding 
definition of an operator that measures $E_{ADM}$  (which is given 
here).   These problems occur in Thiemann's new formulation of quantum 
gravity. Related issues arise in some other approaches such as that
of Borissov, Rovelli and Smolin.  A   new approach to the Hamiltonian 
constraint, which may avoid the  problem of the lack of long  ranged 
correlations, is proposed.

\blankline
${}^*$ smolin@phys.psu.edu
\eject

\section{Introduction}

A non-perturbative quantum theory of gravity must satisfy
at least two criteria to be a candidate for a description of
nature: it must be well defined as a quantum theory and
it must have a good classical limit that guarantees that
the classical Einstein's equations are satisfied  
approximately in an appropriately defined classical regime.
This second criteria is necessary because a non-perturbative theory
is not going to be defined in terms of classical fields, but in
terms of objects defined without reference to a classical
background geometry.   The classical metric must emerge
in an appropriate limit or regime, and it must then satisfy
the appropriate equations.

The classical limit is certainly a significant issue in the 
family of approaches to quantum gravity that have emerged
based on the joining of the Ashtekar-Sen
formalism\cite{abhay1}, 
or one of its variants\cite{fernando-real,renata-real},
with the notion of states based on 
Wilson loops
\cite{tedlee,pl,lp1,lp2,carlo-review,ls-review,aa-review,BGP,GP,g5}.
In these formalisms, there is no background metric, instead,
diffeomorphism invariant and physical states are described
in terms of topological and combinatorial information encoded
in embeddings of spin networks\cite{roger-spinnet} in 
space\cite{sn1}, up to 
diffeomorphisms of the spatial three 
manifold $\Sigma$.
Further, area and volume are discrete\cite{volume1,ls-review} 
and there
are straightforward  graphical formulas for the action of
dynamical operators such as the 
Hamiltonian 
constraint\cite{ham1,roumen-ham,ham2,renata2,GP2,ezawa}.
These results have turned out to be robust:
they arise when the theory is defined through
several different regularization 
procedures\cite{AL1,L1,renata1,renata},  
they are independent of
coupling to matter 
and they are confirmed by rigorous 
methods\cite{rayner,AI,g5,JB1,AL1,L1}.   

Because the Planck scale description
is based completely on the topology and combinatorics of graphs,
it is to be expected that problems with the classical limit might
arise.  This is because the kind of long ranged behavior 
characteristic of classical general relativity may be only
expected to emerge from a discrete dynamics at a critical
point, when the correlation length diverges.  Indeed,
exactly this is seen in a number of approaches to quantum
gravity based on discrete structures, such as dynamical
triangulations and Regge 
calculus\cite{regge,AM,Ambjorn,reggemodels,review-triangles}.  
One sees in
these cases that there are no continuum
limits that could lead to a recovery of the classical 
physics unless the
bare Newton and cosmological constant 
are tuned to a critical 
point\cite{AM,Ambjorn,reggemodels,review-triangles}.  
This suggests 
that without fine tuning,  non-perturbative approaches
to quantum gravity in which the quantum geometry is
discrete at Planck scales may lack classical limits\cite{cosmocritic}.

Another piece of evidence that the existence of a good
semiclassical limit is sensitive to the Planck scale physics has
emerged from the study of black holes.  Several authors
have studied the hypothesis that the area of any surface,
and thus that of an event horizon
is a discrete quantity\cite{BM,volume1,AL1}.  It turns out that 
whether a theory that predicts a discrete spectra for the area
of the horizon has 
a good classical limit depends on the detaled spectrum.  
If the spectrum is equally spaced, as proposed in \cite{BM},
then there is no good classical limit, because the longest wavelength
radiated is proportional to the Schwarzchild radius of the black hole.    
This means that the continuous Hawking spectrum is not reproduced,
even in the limit of large black holes\cite{lsmatters}.  This is
true, even though the level spacing in the area spectrum proposed
is of the order of the Planck area.  Remarkably, a small
change in the formula for the spectrum, to the one which results
from canonical quantization\cite{volume1}, 
resolves this problem\cite{resolution}.

Recently, in a remarkable series of 
papers\cite{tt-wick,qsdi,qsdii,tt-length,tt-volume,tt-summary}, 
Thomas Thiemann has
proposed and developed a new approach to quantum gravity along
the lines of the non-perturbative canonical approaches, 
which makes it possible to overcome a number of
problems with previous formalisms.  Chief of these
is that it is possible to write closed form expressions for the
action of the Hamiltonian constraint and physical inner product
for the theory with Lorentzian signature, eliminating the  difficulty
known as ``the problem of the reality conditions".  
Thiemann introduces completely regulated forms
of the Hamiltonian constraint for both the Lorentzian and
Euclidean signature case and finds the full set of solutions to the 
constraints for these forms, which are normalizable under
a satisfactorily defined physical inner product.   Other
developments include a length operator with discrete
spectrum\cite{tt-length} and an explicit form of a transformation
between the euclidean and lorentzian form of 
the theory\cite{tt-wick}.
Furthermore,
building on previous work 
with and by collaborators\cite{g5,AL1,L1}, this is
done in the context of a completely rigorous formulation,
so that all these results are presented
as theorems in mathematical quantum field theory.

Whatever follows, Thiemann's papers represent a significant
achievement that contain one possible completion of the
program of constructing quantum general relativity 
non-perturbatively.    The theory presented there may indeed
be considered to be a well defined diffeomorphism invariant
quantum field theory gotten by applying a certain quantization
procedure to general relativity.  This makes it urgent to consider
the second question, that of the classical limit.  
One goal of this paper is to investigate whether Thiemann's
formalism may have a good classical limit.
Unfortunately, we find evidence that, at least as defined so far in 
\cite{qsdi,qsdii},  the theory fails to have a good classical limit.

The reasons for this are developed in the
next two sections.    
I should note
that the detailed discussion is not 
self contained, but is based on the definitions and
results in  \cite{qsdi,qsdii}.  
However, the basic issues are not difficult to describe on an
intuitive level.  The main problem is that generic states lack long
distance correlations of the type expected if the theory has
massless particles.  This problem arises
because Thiemann's definition of the
Hamiltonian constraint acts entirely within disjoint 
regions in a generic spin network
state, and does not convey any information between them.
Furthermore, these regions have finite expectation values of
volume, when measured with the operators and inner
product used in \cite{qsdi,qsdii}.  As a result of this, generic
states have arbitrary finite correlation lengths.    

A  consequence is that the hamiltonian of the theory,
defined on an appropriate space of asymptotically flat states,
is not bounded from below.  As I will show in the next
section, in states that have a good classical limit asymptotically,
the information near infinity that is picked up in the $ADM$ mass
in no way constrains the behavior of the state in the interior,
as one can generically find regions in the interior that
are completely uncorrelated with the regions near infinity.
This means that there are no long ranged effects of the kind that
one sees at work in the proofs of the 
positive energy theorem that prevent initial
data which are asymptotic to negative mass Schwarzchild metrics
from being extended over the whole spatial manifold.  
As a result, one can construct exact normalizable 
solutions to the constraints 
that are eigenstates of a natural form of the 
$ADM$ energy operator with any
sign of the mass.

 This raises several questions, which will be discussed at length in
section 4.  First, are the problems restricted
to this one formalism, or are they more general?  Second, are there
ways that we might modify the quantum theory so as to avoid these
difficulties and insure the existence of a classical limit?  In fact, for
reasons I will discuss, it seems
that the existence of a good classical limit is quite sensitive to the 
form
of the Hamiltonian constraint.   Other forms which have been 
proposed
in the literature, such as \cite{ham1,roumen-ham,ham2} also
have a problem with correlations restricted to bounded regions.
However, it is not difficult
to modify the form of the quantum Hamiltonian constraint so as
to eliminate this difficulty.  One way to do this is proposed
in section 5.   An interesting and provocative fact is that it is
difficult to see how the form proposed there could
be deduced from a conventional point split regularization procedure.
This issue is discussed
in the concluding appendix\footnote{Another question 
that might be asked is whether the theories discussed
here may have continuum limits, even if these do not give general
relativity in four dimensions.  This question is explored in \cite{np-strings}
in which it is shown that a certain sector of the theories has collective
coordinates that can be understood in terms of strings in three spacetime
dimensions.}.

I should also stress that there have been other suggestions
as to how the problems I discuss here may be avioded.
These are discussed in section 4.

Finally, before beginning, I should emphasize that
the following considerations are based partly on physical arguments which
are not at the same level of rigor as the original paper. 
It may be possible to fill in the details to obtain rigorous
results along the lines sketched here, but this has not been done.
I should also warn the reader that 
in the course of the argument a few assumptions are made,
whose justification is that their negation would
imply that the theory lacks a good classical limit.  So the 
argument that follows has partly 
the form of a proof by contradiction, one can challenge
certain of the assumptions, but only at the cost of
admiting from the beginning that the theory cannot
reproduce the physics of general relativity.

\section{An ADM energy operator and its spectrum}

The key issue involved in defining the ADM energy non-
perturbatively is in giving a notion of asymptotic flatness
that may apply to non-perturbative states constructed
from spin networks which are also exact solutions to the
constraints, when those are smeared with lapses and shifts
that satisfy appropriate fall off conditions.  I propose here
one way in which this might be done, which has its origins
in the notion of a weave\cite{weave}, which is a quantum state built
from spin networks that is associated with a classical metric.
One of the strange things about the theory, which is shared
by all of the approaches based on the connection or loop
representation is that there can be states which are at the
same time exact physical states and weaves that approximate
a classical three geometry.  

Of course, the three metric is
generally not a physical operator, as it does not commute
with the Hamiltonian constraint.  Thus, it is unlikely that any 
interpretation of the physics of such states based on the
corespondence to a particular three metric is going to be
meaningful.  One way to understand this is that 
even if one can associate a three
metric with a physical weave state, which is an exact
solution to the constraints, it cannot be meaningful
as one does not know which three surface that metric
is to be associated with.   

However, there is one case in which it may
be meaningful to find a limited 
correspondence between a classical
metric and a physical state.  This is in the case that the
state satisfies also some condition of asymptotical flatness.
In this case, there may be gauge invariant information in
the state-metric correspondence, to leading order in
$1/r$, where $r$ is an appropriate radial coordinate.
Some of this information is coded in the $ADM$ mass.
The basic strategy we follow here is to exploit this fact,
to show that positive energy is violated as easily as it is
satisfied in the quantum theory defined by \cite{qsdi,qsdii}.

\subsection{Preliminaries concerning weaves}

I will need to discuss primarily weave states associated with
conformally flat metrics of the form 
$q_{ab}=\Omega^2 (x) q^0_{ab}$,
where $q^0_{ab}$ is a flat metric.  
The weaves will be defined with respect to the volume operator,
$\hat{V}[{\cal R}]$ alone, where ${\cal R}$ is any region of
three space $\Sigma$.   It is true classically that $V[{\cal R}]$
for every $\cal R$ determines $\Omega$ and hence $q_{ab}$,
within this category.  We will also assume the existence of
orthonormal coordinates $\hat{x}$ associated to $q^0_{ab}$
that cover $\Sigma$, which will have the topology of $R^3$.

I will discuss two kinds of weaves: eigenweaves and 
expectation value weaves.  

The first will be eigenweaves.  An eigenweave
will be a linear combination of spin network states\cite{sn1} 
$|\Gamma >$, all with support on the same graph.  
A key point to remember in the following is that associated to
vertices with four or more incoming edges are generally finite
dimensional state spaces\cite{sn1}.  Generally bases may
be chosen for these spaces in which the volume
operator is diagonal\cite{dpr} 

To define a weave we need
two length scales $R$ and $L$ such that 
$R >> L >> l_{Pl}$.  We will say that  $\Omega $ is slowly varying on
the scale $R$, if
$\delta^{ab} \partial_a Ln(\Omega) \partial_b Ln(\Omega) < R^{-2}$.
Then for all cubic regions $\cal R$ with classical volume 
greater than $L^3$ we require two conditions on $|\Gamma >$.
The first is that the state be isotropic, when averaged over
regions of size $L$, with respect to the flat metric
$q^0_{ab}$.  This means that no operator that
averages information over regions of size $L$ will be
able to determine a prefered direction around any point
(up to fluctuations of order $l_{Planck}/L$.)
As a result, if the state is a weave state of some
metric, it is one that is conformally related to $q^0_{ab}$.
The second condition is that,
\f
\hat{V}[{\cal R} ]|\Gamma > = |\Gamma >
\left (  {V}[{\cal R} ] (q_{ab}) + O(l_{pl} /L)  
\right ) \label{eigenweave}
\ff
where ${V}[{\cal R} ] (q_{ab}) $ is the classical observable
that measures the volume of the region $\cal R$ as a function
of the metric $q_{ab}$.

An expectation value weave state $|\Psi >$ 
associated to $q_{ab}$ does not necessarily
have support on a single graph.  It must 
be also isotropic, while the second condition is
modified to,
\f
<\Psi | \hat{V}[{\cal R}] |\Psi> =  <\Psi |\Psi >
\left ( {V}[{\cal R} ] (q_{ab}) + O(l_{pl} /L) \right ) .
\label{exweave}
\ff

We may ask whether the states $|\Gamma > $ and
$|\Psi >$ must satisfy similar conditions with respect to other
three metric observables such as areas of large surfaces,
or lengths of appropriately averaged curves.  It is important
to note that generically weave states do not have
this property.  This reflects the fact that in quantum mechanics
functional relationships between classical observables are not
obeyed by expectation values of states.  However, it is
possible to construct weave states that satisfy the appropriate
conditions for both area and volume\cite{carlo-talk}.  We may call
these consistent weave states.  It is 
good that they exist, were it not the case the theory would already 
have a problem with a
classical limit as there will be no 
states in which these different operators satisfy their
classical relationships.   However, this raises a dynamical problem,
which is to understand why the ground state of the theory should
be consistent in this sense.  In this paper we will assume that all 
weave
states are constructed to be consistent.

We may then construct a consistent weave state
$|\Gamma_0 > $ that is an eigenweave of the flat metric $q_{ab}^0$.  
We will
construct it from identical vertices each with $m$ incident
edges.  $m$ might be $6$, as in the case of a cubic lattice,
but will be required to be at least $4$, so that each node
may contribute to the volume operator.
By the first
condition it is required to be isotropic.
We will also require that, when averaged averaging over the scale $L$,
the expectation values of $\hat{V}[{\cal R}]$ will 
be invariant, , under the Euclidean group of 
$q_{ab}^0$, up to first order in 
$l_{Planck}/L$.   It follows from symmetry that 
any state that satisfies this must be a weave
for a flat metric 
proportional to $q_{ab}^0$ when measured
by any other averaged observable.  We will then adjust the
graph and its labelings so as to have a consistent eigenweave
of $q_{ab}^0$.  

For simplicity we will assume that all the edges
have the same spin $j_0$.     The condition of average isotropy
can then be met if the nodes are scattered randomly
in space (as determined by $q^0_{ab}$), with each
one connected to the $m$ closest, and if the spins
on the edges are small so the volume contributed by
each node is order the Planck volume.

Associated to the vertices with $m$
edges of spin $j_0$ there is a finite dimensional Hilbert
space ${\cal H}_{j_0}$.  $\hat{V}$ induces in each of these
a $p$ dimensional Hermitian\cite{dpr} matrix $\hat{v}_{m,j_0}$.
We will assume that this is non-degenerate (if it is not the
following argument must be altered accordingly).
It will then have eigenvalues $w_1,...w_p$ ordered
in terms of size.  Their normalization is chosen 
so that eigenvalues of the volume are $l_{pl}^3w_i$,
$i=1,...,p$.  We may note that  
the $w_i$ are of order one for small
spins\cite{renata,dpr,tt-volume}.  
For the following we will need to define $w^\prime =(w_1+w_p)/2$.

We then construct the state $|\Gamma_0>$ as follows.
In each cubic box of size $L$ we distribute
randomly $N$ nodes and connect
them up into a network with edges with spin $j$
as we described so that
\f
N={1 \over w^\prime }{L^3 \over l_{pl}^3}
\ff 
We choose randomly half the nodes and give them
the highest eigenvalue $w_p$, while we give the
other half the lowest, $w_1$.
It is straightforward to check that this satisfies the
conditions of both kinds of weaves.

Now we want to construct weaves for the conformally flat
metrics $q_{ab}= \Omega^2 (x) q_{ab}^0$,
with $\Omega$ slowly varying.  This is easy to do
if we restrict ourselves to cases in which
\f
{w_1 \over w^\prime } < \Omega^3 < {w_p \over w^\prime} 
\label{limits}
\ff
which will be sufficient for our purposes.  We do this by keeping
the same graph of $\Gamma_0$ and varying only the 
distribution of the two eigenstates
of volume associated to $w_1$ and $w_p$.  
We do this in a way that preserves the isotropy of the
states, to order $l_{Planck}/L$, when averaged over 
regions of size $L$.  This implies that the resulting metric
must be conformally flat.

It is then clear we can realize
\ref{eigenweave} or \ref{exweave} for 
every box of size $L$ if we choose a distribution of
eigenvalues on the nodes in each box 
such that
\f
{\bar{w} (box) \over w^\prime } = \Omega^3 (x_{box})
\ff
where $\bar{w} (box) $ is the  average value
of the nodes in the box and   $x_{box}$ is its center.
Let us pick one such state and call it $|\Gamma_0^\Omega >$.

Now, let us consider a class of spherically symmetric metrics
built the following way.  Pick two radii $R_1 < R_2$  
much larger than $L$.  and such that one can find a mass $M$
(with $GM >> L$) such that 
\f
1+ {GM \over 2 R_2}  < (w_p /w^\prime )^{1/6}
\ff
and 
\f
1- {GM \over 2 R_2}  > (w_1 /w^\prime )^{1/6}
\ff
Then consider two choices of $\Omega$, called
$\Omega^\pm$ given by 
\f
\Omega^\pm = (1 \pm {GM \over 2r} )^2 , \ \ \ r> R_2
\ff
and 
\f
\Omega^\pm = 1 , \ \ \  r< R_1
\ff
with $\Omega^\pm$ each chosen to be some smooth,
slowly varying
function of $r$ for $R_1 < r < R_2$.

Then we have two eigenweave states, which correspond to a three
geometry that is either positive or negative mass Schwarzchild
externally, each of which are smoothly joined to a flat metric
on the interior.  Let us call these weaves 
spin network states $| \Gamma_0^\pm> $.  Note
that by making $R_1$ and $R_2$ large enough we can
construct these states for any $M$, positive or negative.

We may note that all of the weave states we have been
discussing are solutions of Type I (by the definition of
Theorem 1.1 of \cite{qsdii}) to the Hamiltonian 
constraint.  This is true for both the Euclidean and
Minkowskian operators.  Furthermore, for each state
we have been discussing a diffeomorphism invaraint state
can be given which is the characteristic state of the
corresponding diffeomorphism class\cite{sn1,g5}.

Thus, we see that there are physical, normalizable weave
states associated to three metrics which are asymptotically
Schwarzchild, for any values of the mass, both positive 
and negative.  However this is not yet enough to conclude
that the theory is unstable, as we cannot directly associate
the three geometry to the state, as we discussed above.
The question is whether we can define a gauge invariant
operator which measures the $ADM$ mass in a  
class of states restricted by some suitable definition of
asymptotic flatness, and whether this can be done in 
such a way that the $ADM$ mass extracted is in fact
the mass associated with the asymptotic behavior of
the metric of the weave.  This is done in the next
subsection.  After this, the results are extended to the
generic solutions of type II.

\subsection{Quantum asymptotic flatness and the ADM operator}

The main problem to define an operator for the ADM energy
in non-perturbative quantum gravity is to give a 
definition of asymptotic flatness appropriate for the quantum
theory.  Here I will not attempt to be rigorous, but just to
give the main idea which is necessary to introduce an
operator that measures the ADM energy of a quantum state.

Let us first recall that an asymptotically flat three metric is
defined with respect to some given flat background metric,
$q_{ab}^0$, which can be written in appropriate Euclidean
coordinates, $\hat{x}^{\hat{a}}$ as $\delta_{\hat{a}\hat{b}}$.
We need also a choice of a radial coordinate $r$ defined
with respect to $q_{ab}^0$, and the associated
angular coordinates $\theta$ and $\phi$.  
Given these a metric $q_{ab}$ will be
considered asymptotically flat (in isotropic coordinates) 
if it take the form in the
coordinates $\hat{x}^{\hat{a}}$:
\f
q_{\hat{a}\hat{b}}= (1+{GM(\theta ,\phi ) \over 2r} )^4 
\delta_{\hat{a}\hat{b}}
+ O(1/r^2 )
\ff
where $M(\theta ,\phi )$ is an angle dependent mass.  
We may note that for the Schwarzchild solution in standard
isotropic coordinates, $M$ is a constant and is equal to the
$ADM$ energy.

The complete 
condition of asymptotic
flatness includes also a specification of the fall off behavior
of the extrinsic curvature $k_{ai}(x)$.   This is because
the $ADM$ energy can be negative, even for the
positive mass Schwarzchild solution, if it is defined
on slices whose extrinsic curvatures do not vanish fast enough near
infinity\footnote{I thank Abhay Ashtekar, Don Marolf
and Thomas Thiemann for pointing out this issue\cite{personal}.}.  
In the classical theory there is then a condition like
\f
|k(x)|^2 \equiv q^{ab}k_{ai}k_{bi}(x) =O(1/r^4 )
\label{kasymp}
\ff
as $r \rightarrow \infty$.   Thus we require also a quantum
mechanical version of this condition.  How this is to be
formulated and satisfied is discussed at the end of this section.

The $ADM$ mass is defined by
\f
E_{ADM} \equiv \lim_{r \rightarrow \infty} E(r)
\ff
where 
\f
E(r) \equiv {1 \over 16 \pi G} \int_{S^2(r)} d^2S_c
q^{0ca}q^{0bd} \left ( \partial_d q_{ab} - \partial_c q_{bd}
\right )
\ff
Because any asymptotically flat metric is conformally flat
up to the order measured by $E_{ADM}$, we have
\f
E(r) = {-1 \over 6 \pi G} \int_{S^2(r)} d^2S_c
q^{0ca}\partial_a q^{1/2}
\label{eadm}
\ff
up to terms of order $1/r^2$, where $q=det(q_{ab} )$.
From \ref{eadm} we have of course $E_{ADM}=M$ when it
has no angular dependence.  

The gauge invariance in the asymptotically flat case is
generated by ${\cal H}(N)$ and $D(v)$, where the lapse
$N$ and shift $v^a$ are order $1/r$ as 
$r \rightarrow \infty$, which guarantees the
gauge invariance of $E_{ADM}$.

Now, let us make a definition of a quantum state which
is asymptotically flat with respect to a flat metric
$q_{ab}^0$ and a radial coordinate $r$.  Recall first that
we  may define
a weave with respect to several different functions of the
three metric, the area, volume, $q_{ab}$.  A state $|\Psi >$
will be called area- (or volume-, or $q_{AB}$-) asymptotically 
flat if it is, up to terms of order $1/r^2$,  an expectation
value weave of the area (or volume or $q_{ab}$)
corresponding three metric $q_{ab}$ which is asymptotically
flat.

We will call a weave state $|\Psi >$ 
simply {\it metrically} asymptotically flat if it is
area, volume and $q_{ab}$ asymptotically flat for the
same three metric $q_{ab}$.  One way to accomplish this
is with a weave that is isotropic, up to terms of order
$1/r^2$, as this guarantees that all the averaged observables
will have to agree that it is a weave of a metric that is,
up to terms in $1/r^2$, conformally flat.  One then only
has to make sure that the different observables agree
about the scaling of the $1/r$ piece of the metric.  

We may also make such definitions for weave eigenstates,
in which case the definition using eigenvalues 
\ref{eigenweave} replaces
the one using expectation values.  We will call a state 
that satisfies the same definition for eigenweaves,
an eigen-asymptotically flat state.

In both cases, a complete definition of asymptotic flatness will
involve a fall off condition of the expectation value of the
extrinsic curvature, as I discuss below.

Given \ref{eadm} we may define an 
operator corresponding to the
$ADM$ mass, appropriate to 
$q^0_{ab}$ and $r$ as follows.   We define
\f
\hat{E}_{ADM} \equiv \lim_{r \rightarrow \infty} \hat{E}(r)
\ff
where $\hat{E}(r)$ is defined from \ref{eadm} using the volume
operator.
\f
\hat{E}(r)= {-1 \over 6 \pi GL^3 } r^2 \int_{S^2 (r) }
\partial_r \hat{V}(Box(r)^L)
\ff
where $\hat{V}(Box(r)^L)$ is the volume operator on any
region which is a cube of volume $L^3$ centered at a radial
coordinate $r$, with respect to $q^0_{ab}$ and $r$.

It follows from what we have said that 
$<\Psi |\hat{E}_{ADM}|\Psi >$
is gauge invariant with respect to 
${\cal H}(N)$ and ${\cal D}(v)$, with 
$N$ and $v$ bounded by $1/r$, where $|\Psi >$ is a 
volume-asymptotically flat states.  Further, under the
same conditions $\hat{E}_{ADM}$
commutes with ${\cal H}(N)$ and ${\cal D}(v)$ when
acting on the space of eigen-volume-asymptotically flat
states.

We might worry that the operator for $ADM$ energy has
built into it a length scale $L$.  This might be eliminated by
a more sophisticated definition of the operator, however, the
present one is sufficient for our purposes as we are discussing
weave states associated to classical metrics.  In this case it
is acceptable to use a notion of course graining in the definition
of the $ADM$ energy.   Any definition that disagreed with this
one on such states would lead itself to problems with the classical
limit, as it would lead to a disagreement between the $ADM$ energy
and the asymptotic form of the metric extracted by taking
expectation values of appropriate metric observables.

It is now straightforward to show that the spectrum of
$\hat{E}_{ADM}$ is unbounded from either above or below
on the space of solutions to Thiemann's constraints defined
by Theorem 1.1 of \cite{qsdii}.    For we may note
that both of the states $| \Gamma_0^\pm> $ defined
above are eigen-volume-asymptotically flat.  Furthermore,
it follows directly that they are eigenstates of
$\hat{E}_{ADM}$ with
\f
\hat{E}_{ADM} | \Gamma_0^\pm> = \pm M | \Gamma_0^\pm>
\ff

\subsection{Lack of positive energy on Type II states}

One might wonder if this is just a problem for the solutions called
Type I in Theorem 1.1 of QSD II.  If this were so we might worry 
less, as these are solutions which are also eigenstates of the
volume operator, which are thus very special.   For example, while
they are normalizable with respect to the inner product used
in \cite{qsdi,qsdii} one might worry that that is not correct, and that
there is another, physical inner product with respect to which they
are unphysical.
Thus, it is of interest to know how general are the existence of states
with both both positive and negative expectation value of the
$ADM$ mass operator.  Unfortunately, there are much more general
forms for such states, as we will now see.

As a first step we may note that
all that is really required to get the unboundedness of
the spectrum is that the state is Type I (that is
constructed from eigen-vertices of the volume) up to
terms of order $1/r^2$.  
One can contaminate the state with dressed vertices 
of Type II, which are not eigenstates of the volume
operator in their neighborhoods, as long as the 
contribution to the action of the volumes of large
boxes falls off as $1/r^2$.

However, what if we require that the state has a generic
form also out to infinity?  It might be reasonable to require
this if there were a superselection principle of some kind
that ruled out type I vertices.  In this case, however, one
can show that the situation is no better.  Now, we will no
longer have eigenstates of $\hat{E}_{ADM}$, at least
generically.  But it is still possible to find an infinite
number of states $|\Psi >$, all of whose vertices are of Type II,
such that $<\Psi |\hat{E}_{ADM} |\Psi >$ is negative.  
This follows directly from the fact that each node of a
non-extraordinary graph may be dressed completely 
independently of the others.

To see this, let us consider for the moment a small network
with just one ordinary node of the type we
have been considering with its $m$ spin $j$ lines sticking
out of it.  
We may call it $\Gamma_{\hat{n}}$ 
(These are not spin networks, but we can consider
such non-gauge invariant states in this formalism.)  We will 
consider as before
just two of the volume eigenstates at  the nodes,
corresponding to the highest and lowest eigenvalue
$w_p$ and $w_1$.  Let us say that a node in one of these states
is in the state $+$ or $-$.    Let us choose a dressing of each
of these, corresponding to a solution of type $II$ involving
$n$ extraordinary edges added to the one node.  
($n$ may be chosen as we like,
as there are generically solutions for each $n$.)
We then have two states
\f
|\Psi^\pm> = \sum_I |\Gamma_I^\pm> c_I^\pm
\label{dressing}
\ff
where the spinnets $|\Gamma_I^\pm >$ include adding $n$
extraordinary edges to the one node of $\Gamma_{\hat{n}}^\pm$.
There are two such states, distinguished by $\pm$ which
are dressings of the two eigenstates at the node.

Let us assume we may choose them such that 
\f
<\Psi^\pm |\Psi^{\pm^\prime} > = \delta_{\pm \pm^\prime}
\ff
Each of these states is no longer an eigenstate of
volume.  However, each contributes a value to the
expectation value of volume:
\f
v( \pm  )=<\Psi^\pm| \hat{V} |\Psi^\pm >
\ff
In the following we will assume $v(+) > v(-)$.  (If not
we will switch the designations so this is true.)

Now, let us go back to our initial network $|\Gamma_0>$
for the flat metric $\delta_{ab}$.  We will construct a large
set of physical states $|\Psi ( \epsilon_\mu ) >$.  Here the nodes
are labeled by $\mu$ and each of them is put first in the state
$+$ or $-$, which will be labeled by $\epsilon_\mu = \pm$,
respectively.  We then dress each node according to the
prescription \ref{dressing}.  Because the 
Hamiltonian 
constraint factors
into a sum that each acts just around each non-extraordinary
vertex, all of these are physical states of type II.
Now among these are many which are expectation value
weaves corresponding to slowly varying conformally flat metrics.
Further, to any such metric we may associate many such states.

To see this, let us note that for a large region ${\cal R}$ 
\f
< \Psi ( \epsilon_\mu )| \hat{V}[{\cal R}] | \Psi ( \epsilon_\mu )>
= \sum_{\mu \in {\cal R}}  v (\epsilon_\mu ) 
\ff
Given a slowly varying $\Omega$ such that
\f
{\Omega^3 (max ) \over \Omega^3 (min) } = {v(+) \over v(-) }
\ff
we can construct an associated expectation value weave by
distributing the $+$ and $-$ dressings so that for each
box of volume $L^3$ in the flat background metric
\f
<\Psi ( \epsilon_\mu ) |{ V(box) \over L^3 } | \Psi ( \epsilon_\mu )> 
= {\bar{v} \over w^\prime l_{Planck}^3} = 
\Omega^3 (x_{box})
\ff
where $\bar{v}$ is the average of the $v(\pm )$ values over
the dressed nodes in the box.

We then can construct expectation value weaves of this kind
that match the two conformal factors $\Omega^\pm$ of
which are asymptotically positive or negative mass
Schwarzchild, for any mass $M$.   
Let us call these states $|\Psi (\Omega^\pm )>$.  We then
have,
\f
<\Psi (\Omega^\pm ) | \hat{E}_{ADM}  
|\Psi (\Omega^\pm )>= \pm M  
\ff
 
Thus, we have shown in this section that
a well defined operator that measures the $ADM$ energy
has a spectrum that is unbounded from above and below,
and further, even when restricting to pure Type II states,
its expectation value is still unbounded above and below.
This means that the classical limit must fail, in the sense
that there are an infinite number of states that have a good
$ADM$ energy, but whose energy is arbitrarily negative.
We may note that as the $ADM$ energy is gauge invariant,
we cannot be in doubt as to the interpretation of this
result, as we might if it were only a result about the existence
of weave states that are associated with negative mass
Schwarzchild.  

We may note also that it would not help to try to define
another operator that measures $ADM$ energy based on
measurements of other operators such as areas or
lengths or $q_{ab}$.  This is because, if there is to be a good
classical limit it must be that there are isotropic weave states
that are weaves of the same conformally flat metric, which
ever operator is used in the construction of the weave.  If this
were not the case we could not believe that the theory
adequately reproduced the classical three geometry.  
Furthermore, we require only that the states have a good
classical limit to leading order in $1/r$, which is enough
to define eigenstates of the $ADM$ energy.   We have
no need to require that the asymptotically flat states have
a good classical limit in the interior, or to higher order in
$1/r$.
But then, 
assuming only that there are such states,
we see that it is
as easy to construct those that are asymptotically negative
mass Schwarzchild as it is to construct those that are
asymptotically positive mass Schwarzchild, when the definition
given here is used.  But then, as the $ADM$ energy measures
a property of these states that is gauge invariant, there can
be no escaping the conclusion that this means that the 
$ADM$ energy is unbounded in the quantum theory 
defined by \cite{qsdi,qsdii}.

\subsection{Implementing fall off conditions on the extrinsic 
curvature}

One might still worry that the problem is that a fall off condition
for the extrinsic curvature has not yet been imposed.  As this is
necessary for the classical positive energy theorem, it might be
that implementation of this condition in the quantum theory will 
restrict the states that have only positive values of the $ADM$ 
energy\cite{personal}.
To be completely sure that the theory has
an unbounded Hamiltonian one must make sure that the 
states are restricted by a quantum analogue of the fall
off condition for the extrinsic curvature.
The simplest possibility 
is to define define an appropriately
ordered and averaged operator for $k^a_b $,
\f
O^a_b(R) =\int_{\cal R} Tr(\tilde{E}^ak_b )
\label{qk2}
\ff
We may then require that 
\f
<\Psi |  O^a_b ({\cal R})  |\Psi >=O(1/r^2 )
\label{qkasymp}
\ff
for regions of size $L^3$ in the background metric
$q^0_{ab}$ as $r \rightarrow \infty$.
 (We may note that this may be superior to measuring expectation
values of quadratic operators that might represent 
$|k|^2$, as these will have zero point fluctuations that will have to
be subtracted out to define the asymptotic behavior. 

It is, however difficult to imagine that it is not possible to satisfy
an additional asymptotic condition such as \ref{qkasymp}, given
the enormous freedom to construct states which satisfy the
asymptotic conditions for the expectation values of the
metric observables.  For there is no reason not to construct
weave states using more than the two eigenstates of volume
at each node I used above.  Let us suppose we instead construct 
our weaves
using identical nodes that have a large number, $r$ of
possible volume eigenstates.  There will then be many 
ways to get any desired $ \Omega( r, \theta,\phi ) $ 
that satisfies \ref{limits} by mixing the 
$r$ eigenstates appropriately
in each region of size $L$. This is true for either the eigenweaves
or expectation value weaves.  In each region I require
that $\omega^3 = \sum_{i=1}^r v_i n_i $ add up to some 
required number, proportional to $(1+ M(\theta ,\phi ) /2r )^6$ 
where
$v_i$ is the contribution to the volume of node eigenstate $i$
in Planck units and $n_i$ is the number of the $N$ nodes in
the region  that are of this type.
Thus we must satisfy  two conditions, $\sum_i n_i =N$, and that
the sum for $\Omega $ is fixed.  Given $r$ kinds of
nodes there are then an $r-2$ dimensional set of
possibilities of achieving the required dependence on
the conformal factor, for any desired $M(\theta, \phi )$.
It is difficult to see how, if $r$ is large enough, it will not be
possible to use this freedom to match any desired fall off
conditions for the extrinsic curvature such as \ref{qkasymp}.

For example,  
we may  note that by the isotropy of the weave, 
we must have 
\f
<\Psi |  O^a_b ({\cal R})  |\Psi > = C({\cal R}) \delta^a_b
\ff
Given that we have a large number of solutions to the 
problem of matching the fall off conditions for the volume, it
should be trivial to choose among this set to get any 
desired fall off on the functions $C$ as a function of $r$.  
As the regions ${\cal R}$ each contain many nodes, then we will
have on averageÊ$C({\cal R} ) = \sum_{i=1}^r c_i n_i $ where
the sum is again over the nodes in $\cal R$ and $c_i$ is the
average contribution of a node of type $i$  to $C$.  All that is
needed to have $C({\cal R})=0$, on average, is for the 
$c_i$ to span a range of both positive and negative values
such that restricted to the $r-2$ dimensional subspace of solutions
to the metric fall off conditions it is possible always to balence
the positive and the negative $c_i$'s in each region.

To complete the argument the $c_i$'s should be computed,
showing that there are nodes with the required property.
This has not yet been done.   However, it is clear that there is
no reason to expect that the imposition of one further condition
on the nodes of the weave states could save the positivity of
the energy.  The real problem, as I will now describe, is that
we have complete freedom to choose the forms of the solutions
independently in the neighborhood of each node.  This means
that there is no long ranged order in this quantum theory,
of the kind that is imposed by the classical field equations.

\section{The absence of long ranged correlations}

We know in general terms, from the renormalization
group as well as from analysis of many different kinds of
systems, what must be the case if a discrete quantum
system is to have a classical limit described by a classical
field theory with massless quanta.  There must be a
critical point at which the correlation length 
diverges\footnote{I refer here to the application of the
renormalization group to random surface theory and 
second order phase transitions, rather than to the problem
of renormalization in conventional quantum field theories.
The latter does not apply to non-perturbative formulations
of quantum gravity in which diffeomorphism invariance
ensures the finiteness of physical operators\cite{ls-review}
but the former definitely applies.}.
For this to be the case it must be true for generic physical states that
small perturbations made in one localized region can be detected
by measuring some appropriate operator arbitrarily far from the
region in which the disturbance is made. 

I will show here that this is not realized in Thiemann's
space of states described by Theorem 1.1 of \cite{qsdii}.
Instead, a generic physical state describes a quantum
geometry that may be decomposed into regions whose
volumes, defined by taking expectation value with the
physical inner product, are finite, but 
each of which is completely uncorrelated with the others.
This may be shown  by constructing an infinite number of
diffeomorphism invariant, physical Hermitian operators,
that by construction describe local excitations, but which
all commute with each other.  
This means that disturbances do not propagate more than a fixed
finite distance in any given state of this type.  But as the
general solution to the constraints found by Thiemann have
this property, that theory cannot
have a perturbation theory in which massless fields propagate, 
nor may it have a classical limit which is
general relativity.  

The general solutions to all forms of Thiemann's constraints may be
described in the following way\cite{qsdii}.  
We begin with the
space $W_0$ of non-extraordinary spin networks (or more
precisely spin networks all of whose extraordinary edges
carry spin greater than $1/2$.) Among the nodes of such
a vertex are a special class,
called simple nodes, which are those
that may get dressed by the action of the Hamiltonian
constraint.  These are nodes 
with  at least three incident  edges
with spins, $j_i$, $i=1,...,n$ in some arbitrary ordering, 
with generally an additional label at the vertex $r$, such
that the tangent vectors of the incident edges span the
tangent space at the node.  We will label these node
$\hat{\cal N}_{n,j_i,r}$, where we include also implicitly
in the labeling diffeomorphism invariant information about
the linear relations among the tangent vectors of the $n$
edges.  These include all the nodes that  may contribute to
the volume operator, with the definition of volume
used in \cite{qsdi,qsdii}.  
We will also assume that the labels at the vertex
$r$ include the volume plus any additional labels required
to break degeneracy.

The simple nodes of a non-extraordinary spin network may
then be dressed to yield the general solution to the constraints
described in Theorem 1.1 of \cite{qsdii}.  The nodes are
dressed by taking linear combinations of the original graph
with those in which a finite number of extraordinary edges
decorate the region around each node, each joining two vertices
coming out of each simple node.  The result is something like
a spider web around each simple node.  The  
source of the trouble is that each node is dressed completely
independently of the others.

To see this, begin with a non-extraordinary spin network
and proceed as follows.
First, note that each such state is in fact a solution to the constraints
of type I according to Theorem 1.1 of \cite{qsdii}.  Then, to each 
representative of the diffeomorphism class of each such spinnet,
$\Gamma$,
construct a  partition of
unity $N_\alpha$, where  each $N_\alpha$ has support on
a set ${\cal U}_\alpha$ that only includes one of its nodes,
labeled by $\alpha$.   Now let us consider only the
hamiltonian constraint $\hat{\cal H}(N_\alpha )$, which
by construction acts in a neighborhood of only one of the
$\alpha$'th node.  There
are an infinite number of solutions of type II to 
\f
\hat{\cal H}(N_\alpha ) |\Psi > =0
\label{dynamics}
\ff
which may be
constructed from linear combinations of states that
have support on spinnets constructed from the node
$\hat{\cal N}_{n,j_i,r}$  by dressing the edges
incident to it with a finite, but arbitrary number
of extraordinary vertices\cite{qsdii}.  
Each of these may be seen as a linear combination
of open graphs with $n$ external lines, with spins $j_i$,
as the dressing procedure never changes the spins of
the portion of the edge furthest from the original
simple node.
Let us give
a name to such a linear combination, associated with
a linear combination of open graphs a dressed node,
and denote it $\hat{\cal DN}_{n,j_i,r,I}$, where $I$
labels the solution to \ref{dynamics} 
gotten by dressing the nodes.

Note that it follows from Theorem 1.1 of \cite{qsdii}
that there are an infinite number of such solutions
to \ref{dynamics} for each simple 
node $\hat{\cal N}_{n,j_i,r}$. 
Further, we can construct solutions to the all the 
Hamiltonian constraints by dressing in this manner
 all of the simple nodes of a non-extraordinary spinnet
$\Gamma_0$.  If we order the simple nodes arbitrarily
by $\alpha =1,...M$, for a graph with $M$ simple nodes,
we may call such a state $|\Gamma_0 , I_\alpha >$,
where the $I_\alpha$'s label the dressings of the
simple nodes of $\Gamma_0$.  We will assume in what
follows that each possible $I_\alpha$ can be coded
as a real number.  
We may note that
the  edges which connect each dressed node to the rest
of the graph have, in each term in the sum making up the
state, the
same spin as was incident on the corresponding
simple node of $\Gamma_0$.  We call these the 
external edges of the dressed node.

Now we may construct an infinite number of local physical
operators as follows.  Let us construct a linear operator
$\hat{\cal F}_{n,j_i}$ associated to a set of $n$ edges
with spins $j_i$, $i=1,...,n$ as follows.  Acting on a spin
network state $\Psi$ it extracts any dressed node
with $n$ external lines labeled by the set of
spins $j_i$ and produces the state in
which the $n$ lines of that node are tied up in a planar
vertex, so that the result is a closed spin network.   
We
may note that such a node will be possible, by conservation
of angular momentum. By
linearity, it extracts linear combinations such as those
that solved the Hamiltonian constraint we called 
${\cal DN}_{n,j_i,I}$.  We will call the resulting state
with the ends tied up at a planar node ${\cal EN}_{n,j_i,I}$
for the extracted node.

The operator must test to make sure it has cut the state just outside
of a single 
dressed node, which means it must test that the state formed
by cutting is itself a solution to the Hamiltonian constraint.

 If there is more than one such
dressed node in a spin network state 
$\Psi$ the operator $\hat{\cal F}_{n,j_i}$
produces a spin network state which  is a disjoint and
unlinked union of the tied up spin networks produced for
each one.    For example if there are two nodes
with the same external edges, ${\cal DN}_{n,j_i,I_1}$
and ${\cal DN}_{n,j_i,I_2}$ in a state $|\Psi>$ then
$\hat{\cal F}_{n,j_i}|\Psi > = {\cal DN}_{n,j_i,I_1} \cup 
{\cal DN}_{n,j_i,I_2}$.

These operators act to isolate regions of the quantum geometry
consisting of one dressed simple node labeled by  a given
set of external spins.  Furthermore, generically the 
extracted state has finite expectation value of volume.  
For this reason the  operators $\hat{\cal F}_{n,j_i}$
may be considered
local operators.  It is also clear by construction that they commute
with the Hamiltonian constraint, on the kernel, in the sense that
they take solutions to the Hamiltonian constraint to solutions
of the Hamiltonian constraint.  

It is clear that these operators all commute with each other.
However their action is too abrupt, each eliminates all
but one kind of region from the quantum geometry.
But given that these operators exist we can make others
that do more interesting things.  For example,
given any two dressings $I_1$ and $I_2$ of the
simple node ${\cal N}_{n,j_i}$ we can construct
a change operator ${\cal C}_{n,j_i ; I_1 \rightarrow I_2}$
as follows.  Acting on a spin network state $\Psi$, this
locates all instances of the dressed nodes ${\cal DN}_{n,j_i,I_1}$
and replaces them by the dressed nodes ${\cal DN}_{n,j_i,I_2}$.
It is clear by construction that these are physical diffeomorphism
invariant operators, which also have a local meaning in space.
What they do is to search for local regions of the quantum geometry
characterized by certain gauge invariant and 
local properties and modify them locally.
 
It makes sense to say that the properties of a given dressed node
are local because, as discussed in the previous section, such a
dressed node ${\cal DN}_{n,j_i,I_1}$ generally has an expectation
value of volume which is finite.

But it follows trivially that for any two of these change-dressings  
operators
\f
[ {\cal C}_{n,j_i ; I_1 \rightarrow I_2} , 
{\cal C}_{n^\prime ,j_i^\prime ; I_1^\prime  \rightarrow I_2^\prime } 
]=0
\label{commutes}
\ff
whenever $n \neq n^\prime$.

We then have acting on the space of solutions an infinite number
of operators whose construction shows that they act locally
in space, but which all commute with each other.  
Furthermore, the only way such operators can fail to
commute with each other is if they act in regions with the
same external edges, which means they act in regions
whose volumes, defined by the expectation value,
are generically finite.  This shows
that there are an infinite number of degrees of freedom in the
theory that correspond to modes that do not propagate.

This by itself is in conflict with the Einstein's equations, for
which there are no such non-propagating modes.  Furthermore,
as the expectation value of the volume of a dressed node
may be as large as one likes, this is not necessarily a Planck
scale phenomena.  According to \ref{commutes} 
one may alter the state of
the quantum geometry on a region arbitrarily large in Planck
units, in a way that does not propagate to any other regions.

It might be objected that we do not know the classical
interpretation of the operators ${\cal C}_{n,j_i ; I_1 \rightarrow I_2}$
we have constructed.  This is true, but there are classical analogues
of such observables, they correspond to the relational observables
described in many places\cite{relational}, 
in which one first locates a point of spacetime
by making measurements of certain 
fields and derivatives and then, having  
individuated physically
a point or an event, 
measures other fields there.   By construction, such classical
observables are physical, as they are spacetime diffeomorphism
invariant.  To construct their explicit operator representations
exactly is probably impossible, as they require the integration
of Einstein's equations.  

It is clear that the operators ${\cal C}_{n,j_i ; I_1 \rightarrow I_2}$
are quantum mechanical analogues of these relational
classical physical observables.  While it will be difficult to construct
the exact correspondences between them and the  
classical relational observables, we can discover the physical
effects of some of them directly.  For example, among them are
those that act on vertices which are eigenstates of the volume
and replace them by other volume eigenstates.  More generally
the ${\cal C}_{n,j_i ; I_1 \rightarrow I_2}$ will change the
expectation value of the volume in the region around the
dressed node in the quantum geometry.  We may deduce
from \ref{commutes} that one can
make an infinite number of transitions between physical states
that each change the expectation value of the volume in a restricted
region of space,  in which the transitions between physically
distinct regions are completely uncorrelated with each other,
for all time.
This certainly would not be possible in any solution to
Einstein's equations.

\section{Are there ways out?}

Are these problems just pathologies of the particular construction
of the states in \cite{qsdi,qsdii}, or do they reflect more general
problems with approaches to quantum general relativity based
on spin networks states?  The purpose of this section is to
discuss several possible answers to this question.

\subsection{First remarks}

We may first note several things that might at first be thought
to be relevant, but which on exampination have nothing to do with
the issues we are describing here.  

First, the problem is not degenerate states or
the fact that the Ashtekar constraint is of density weight two,
(as was the case with Varadarajan's examples 
of negative energy solutions to the constraints 
described in \cite{madhavan})
because  Thiemann's construction is explicitly designed to 
eliminate these issues.  

Second, there is nothing wrong with the new identities discovered
by Thiemann and exploited in his construction of the Hamiltonian
constraints of the various theories.  These represent important
new insights into quantum general relativity that likely have
many applications beyond the questions described here.
Nor is there anything wrong with the use of a real connection
variable of the kind Thiemann uses, which has allowed him to
solve the reality condition problem.  We may note also that the
problems discussed apply equally to the Euclidean and Lorentzian
forms of the theory.

Third, the problems are not just restricted to Thiemann's
formalism.  For example, analogous problems arise with
the original forms of the solutions to the constraints
in \cite{tedlee,lp1}, in which the states have support only on
diffeomorphism classes of spin-networks made out of
closed loops, without nodes.  These have no volume, but they
can be made to correspond to classical metrics by definitions
of weaves based on area or surfaces or norms of 
fields\cite{weave,ls-review}.
As such a class of such physical states could be defined that
was also asymptotically flat, that included states that 
correspond to three metrics that are asymptotically Schwarzchild,
for both positive and negative values of the mass.

Finally, the issue has nothing to do with the difference
between the connection and loop representations, just as the
lack of boundedness of the energy of the upside-down-harmonic
oscillator has nothing to do with a choice of position or momentum
representation.    While the rigorous methods used by Thiemann
are couched in the connection representation, what is at
issue has nothing to do with subtleties of mathematical quantum
field theory.  It is also easily discussed in the loop 
representation\footnote{This is true generally, any calculation
in quantum gravity so far done in one representation may easily
be done in the other\cite{roberto}}.
Indeed, one has only to notice that Thiemann's basic identities
work equally well in the loop representation so that
Thiemann's forms of the constraint operators may be
constructed directly there.   To see how to do this,  note
that in the loop representation one may also consider
non-gauge invariant states corresponding to open
lines\cite{carlo-talk}.
Consider then an open or closed loop
$\gamma$ based at $x$, with no kink or vertex there 
and a spherical region ${\cal R}^\delta$ around $x$
of radius $\delta$ in some
flat background metric $q^0_{ab}$.  Then it follows right away that
in the connection representation,
\begin{eqnarray}
Tr  ( e_a (x) \dot{\gamma}^a (0) U_\gamma  ) &=&
Tr  ( [ A_a(x) \dot{\gamma}^a (0) , \hat{V} ] U_\gamma  )
\nonumber \\
&=& lim_{\delta \rightarrow 0} 
{1 \over \delta }  [ T[\gamma ] , \hat{V}[{\cal R}^\delta ] ]
\label{yes}
\end{eqnarray}
Thus, as long as one is interested in the end only in using
Thiemann's identities in cases in which
the index on $e_a (x)$ is tied up with the tangent vector of a curve
one can use always expressions  such as these to define
an extension of the loop representation.

To do this one follows 
the philosophy of its original construction
of the loop representation in quantum gravity\cite{lp1}.
We {\it define} an operator directly in the
loop representation that corresponds to \ref{yes}
\f
\hat{T}_0 [\gamma ] \equiv 
lim_{\delta \rightarrow 0} 
{1 \over \delta } [\hat{T}[\gamma ] , \hat{V}[{\cal R}^\delta ] ] .
\label{def1}
\ff
This defines a new kind of loop representation operator
$T_s [\gamma ]$ which has an insertion of 
$e_a\dot{\gamma}^a(s)$ at the point $s$ of the loop.  By analogy
we may speak of the insertion of $e_a\dot{\gamma}^a(s)$
as the ``foot" of the operator, whose action on loop states
is defined by \ref{def1}.

Thus, the situation is exactly the same as in the case of the
various definitions of volume operators\cite{L1}.  One may define
different regulated operators, but whatever may be done
in the connection representation may be done also in the
loop representation and visa versa.  So the issue
of the representation used may be separated completely 
from the
issue of which regularization procedure is chosen.

This is, of course, not to
say that one formalism may be more useful than the other
for certain purposes.  For example, at the present time, 
rigorous statements may be made more precisely in the connection
representation, while the extension to the $q$-deformed
spin network case can only be done directly in the loop
representation.

Now that we have discussed where the problems do not
lie, let us discuss several possible places the approach
might be modified so as to avoid them.

 \subsection{Some possible loopholes}

1)  It might be that Thiemann's  approach is correct, but that
the space of normalizable physical states described by
Theorem 1.1 of \cite{qsdii} actually consists of more than one
disjoint sector, only one of which is physically relevant.  If this
is the case then there may be a superselection principle which
excludes those states that have the problems we 
have discussed\footnote{Along these lines we may mention that one 
may object that the
weave states are strictly speaking not in the state space
defined in \cite{qsdi,qsdii} as it includes only states constructed
with finite numbers of nodes\cite{personal}.  
However, it is hard to imagine
that there is not a straightforward extension that includes solutions
with countably infinite numbers of nodes.  (Again, the lack of
long ranged correlations makes this seem likely.)  If there is not
then the theory would not be capable of incorporating an
asymptotically flat regime, which would in itself be cause for
worry.}.

It seems, however, that this is not likely on physical grounds.
First, to avoid the problem of the 
lack of long-ranged correlations, the
allowable states would have to be dressed states of only one
simple vertex.  This would greatly restrict the theory and make
weave states based on the volume operator 
impossible, as each physical state will have only one node
in each term in the sum that makes it up that contributes
to volume.  This makes impossible also a notion
of asymptotic flatness using states that are asymptotically
weaves along the lines developed here.  We might
add also that  there is no example
of an interacting quantum field theory with an unbounded 
Hamiltonian
where the boundedness is restored by a superselection
principle.   Still, this is a possibility that should be explored.

2)  It may be that the correct physical inner product differs from the
one based on the Hilbert space used in \cite{qsdi,qsdii},
 in such a way that most of the solutions of both Type I and
Type II 
are unphysical.  Instead, a new physical
inner product might allow only states that resolve these
two issues\cite{personal}.  This would clearly be relevant,
as one cannot show the boundedness of the Hamiltonian
for free quantum field theory, or even the harmonic oscillator,
independently of the inner product.   

In this respect it is interesting to note that the one exact 
physical state that
is known to have a good semiclassical limit\cite{chopinlee}
is the Kodama state \cite{kodama} which is the exponential
of the Chern-Simons invariant of the Ashtekar-Sen connection.
This state does not exist in the state space defined in
\cite{qsdi,qsdii} as it requires a quantum deformation of the
loop representation\cite{sethlee,rsl} based on quantum
spin networks\cite{sn-lou}, such that the deformation parameter
is proportional to the cosmological constant.

3)  If the problem is not in the form of the inner
product than it can only be resolved by changing the dynamics.
In the rest of this paper I will explore the option that the
problems we have discussed may be resolved by modifying
the form of the Hamiltonian constraint operator.  

\subsection{Could changing the definition of the Hamiltonian
constraint help?}

It is clear from the discussion of the second problem that the
issue has at least partly to do with the fact that
Thiemann's Hamiltonian constraints distinguish the two
kinds of vertices, extraordinary and simple.  The result is that
the ``evolution" of any state under the Hamiltonian constraint
divides into regions, each associated with a single simple
vertex which is then dressed with extraordinary vertices.
These different regions do not communicate with each other
under the dynamics generated by the Hamiltonian constraint.
Furthermore, there are an infinite number of exact solutions
for dressing the vertices in which the volume of each of these
regions, defined by the expectation value of the volume operator,
restricted to a region containing only one simple vertex, is
bounded.  Thus, in all of these states there is no propagation
of physical correlations beyond bounded regions.
As it is clear that there is nothing like this in classical
general relativity (even horizons allow one way communication)
this is a feature that will have to be modified if the theory
can have a good classical limit.

One may ask if this might be overcome by making a
different choice of Hamiltonian constraint operator within
Thiemann's basic framework.  For example, it may be that
a symmetric operator may be constructed, not along the lines
done in \cite{qsdi,qsdii}, but by adding to the operator
its hermitian conjugate with respect to the inner product.
This might produce an operator whose solutions were not
decomposable into regions each with finite expectation value
of volume.  

However, it seems likely that this will not eliminate the
problem of the bounded correlations.  
The problem is that if the hermiatian conjugate is taken in
the inner product used in \cite{qsdi,qsdii} than it too is
diagonal in the blocks defined by the graphs which dress
the different simple vertices.  This means that while the
Hermitian conjugate may eliminate extraordinary edges,
it can also not have any terms which connect the states
of distinct simple vertices.  This means that the symmetric
operator is block diagonal as well, and the different sectors
associated with the dressings of different simple vertices
do not get mixed.  

 However, the problem is not only the distinction between
simple and extraordinary nodes.  
It is not hard to find regularization procedures that lead to a 
form of the hamiltonian constraint that is 
is block diagonal in the spin network basis, so that it does not 
propagate
physical information through a whole graph, even if the  
distinction 
between different kinds of vertices is eliminated.  
One of these is the definition of the 
Hamiltonian constraint (and Hamitonian) 
given in \cite{ham1,roumen-ham,ham2}.
That form of the Hamiltonian constraint dresses any vertex,
in which there are non-colinear incident edges, 
bivalent or higher, with two new trivalent vertices,
joined by a new edge with spin $1/2$.  As a result,
vertices that are created by the
Hamiltonian constraint are themselves dressed by other
vertices, in contrast to what happens in Thiemann's
definition.  Furthermore, there is no requirement that only
vertices with three independent tangent vectors are
dressed, so that all trivalent vertices are on an equal
footing.

To find solutions to this form of the constraint one may
deal with the problem of the reality conditions by
defining a Hermitian form of the constraint, which is
\f
{\cal H}(N)={1 \over 2} \left ( {\cal C}(N) + {\cal C}^\dagger (N)
\right ).
\label{hermitian-form}
\ff
The details of this are discussed in \cite{np-strings}, but
the main point is easy to explain.  One begins with
a skeleton consisting of an arbitrary spin network.  One
then dresses each vertex with new trivalent nodes,
each of which is dressed
in turn. In this way associated to every pair of
tangent vectors at a node of the original skelaton an
infinite dimensional
space of states is constructed whose basis elements form 
planar and fractal structures.   Generic solutions than
require an infinite number of networks.
However, the solutions may still be found independently
for each region which dresses a node of the original skeleton.

One may ask whether the regions generated have finite
or infinite expectation values of the volume.  This depends on
the form of the volume operator used.  If one uses
the ordinary operator each trivalent node contributes
zero volume \cite{renata,renata1,gi,dpr,rsl}.  
But this is remedied if the formalism
is deformed to the quantum spin network 
case\cite{sethlee}, in which
case all trivalent vertices contribute to the volume\cite{rsl}.  
One can see that in this formalism the volume of
the region affected by the initial conditions at an initial
vertex grows generically with repeated actions
of the Hamiltonian constraint\cite{rs-personal}.   
This means that
after an infinite number of iterations of the Hamiltonian
constraint, which is necessary to produce a solution, generic states
have long ranged correlations.

At the same time, the correlations are still restricted to the regions
that are associated to each node of the original skeleton.  This
seems a serious problem, whether or not the regions have
finite expectation value of volume.
To avoid this one must use a definition of the Hamiltonian
constraint that does not have the feature that solutions are
generated by dressing skeletons.  One way to do this
is presented in the following.  

\section{A new proposal for the regulated Hamiltonian
constraint}

Up till now most approaches to the regularization of the Hamiltonian
constraint have followed followed a common 
methodology\cite{lp1,ls-review},
which is to {\it construct} the operator through a regularization
procedure in which the classical expression for the constraint
is expressed as a limit of point-split operators.  However, it
must be emphasized that this
is not necessary, as all that is required is that the operator
have a form that leads to the correct classical limit.  As the
issue of the classical limit is not straightforward, as we have
seen here, we are in the position of having to find a suitable
operator by trial and error.  If a given methodology fails to
produce an operator with a good classical limit, we may widen
the methodology.  One way to do this is to require only that
the operator, written in the connection representation, have
an action which approximates, for slowly varying connections,
one of those produced by an actual regularization procedure,
up to terms small in Planck units.
This is a well understood procedure in ordinary lattice gauge
theory\footnote{We may note that many of the 
regularization procedures so far
proposed in the continuum, including Thieman's \cite{qsdi,qsdii} 
and the one studied by Borissov, Rovelli and the author
\cite{ham1,roumen-ham,ham2} 
have steps which require
additional operator dependence, besides those which are 
expressed in terms of the canonical variables.  This 
implicit dependence arises 
in the way that the loops used in the regulated
operators are chosen to conform to certain features of the
geometry of the spin-networks that parameterize the spin network
basis.  Such operators exist naturally 
in the loop representation, as shown by the example of the
operator $\dot{\gamma}^a(x)$  defined by
$\dot{\gamma}^a (x) |\Gamma > = \int ds \dot{\gamma}^a (s)
\delta^3 (x ,\gamma (s))|\Gamma >$.  But it
is not know whether they
can themselves be constructed through any regularization procedure
from functions of the canonical variables.  For this reason it is not
known if they can be constructed in the connection representation
formulation employed in \cite{qsdi,qsdii}, if one requires a
construction that uses only   operators that
represent the classical canonical variables.   
Of course, this is consistent with the philosophy just enunciated
for the regularization of the Hamiltonian constraint.  The
operator I am about to introduce takes advantage as well of
the freedom to employ operators such as $\dot{\gamma}^a(x)$
in order to make the
regulated operators perform combinatorial operations directly
on the spin networks.}.

It is likely that in order that generic states not be decomposible
into uncorrelated regions, the action of the Hamiltonian constraint
must act more freely on the space of spin-networks, which means that
acting on a node $v$ of a spin network, it must alter the edges in
a neighborhood of $v$ that includes also its neighboring 
nodes\footnote{To my knowledge the
only previous proposals for the Hamiltonian constraint
that have this property are those defined on a lattice,
\cite{pl,renata2,ezawa,GP2}.}.  

It is not difficult to invent operators that do this, using Thiemann's length 
operator\cite{tt-length}, as I will now describe.

The action of the new form of the Hamiltonian constraint,
${\cal C}_{new}$, is defined by the following four step 
procedure\footnote{A slightly different version of this operator,
which takes trivalent networks only to trivalent networks
is used in \cite{np-strings}}.

\begin{itemize}

\item{\bf R1} ${\cal C}_{new} (N)$ acts on an element $\Gamma$ of the
spin network basis at each pair of non-colinear
edges $e_1$ and $e_2$ of every node $v$. 
The operator acts on a node $v$ and a pair of its edges
$e_1$ and $e_2$ by modifying a subnetwork that includes
these elements as well as the  
 nodes $v$ is connected to by  $e_1$ and $e_2$,
which will be called $v_1$ and $v_2$.  The subnetwork
to be modified includes as  well the edge connecting $v_1$ and $v_2$
if it exists\footnote{For simplicity we may restrict attention to
graphs in which there is at most one edge connecting any two
nodes.}.

\item{bf R2}  
Suppose that there is an edge joining $v_1$ and 
$v_2$, which will
be called $e_{12}$.  The action of ${\cal C}_{new}$
 produces a sum of six terms in
which the colors along $e_1$, $e_2$ and $e_{12}$, which
we call $i,j$ and $k$ respectively, are updated by $\pm 1$.
Each is multiplied by an amplitude 
$A_{\pm , \pm^\prime , \pm^{\prime \prime}}(i,j,k;r,s,t)$
which I give below.
Here we assume that each of the nodes is written in the form
in which the two edges in the problem are joined to a third
edge at a trivalent vertex with an edge with definite color.
In the case that the node has more than $3$ incident edges,
this new edge is internal to the node.  But, using the
recoupling identities, any higher than trivalent node can
be represented in terms of trivalent nodes joined by
internal edges of length zero\cite{sn1}.  The colors
associated with these edges for $v, v_1$ and $v_2$,
respectively, are $r,s$ and $t$.  
$\pm , \pm^\prime $ and $ \pm^{\prime \prime}$ refer
respectively to the updating of $i,j$ and $k$.  The
amplitude is then,
\begin{eqnarray}
A_{\pm , \pm^\prime , \pm^{\prime \prime}}(i,j,k;r,s,t) &=&
\pm^{\prime \prime} ij  
\left \{ {i i i\pm1}; {112}  \right \}
\left \{   {j j j\pm^\prime 1}; {112}  \right \}
\left \{   {i\pm 1 i r }; {j\pm^\prime 1  j 1 }  \right \}
\nonumber \\
&&\times 
 \left \{   {i\pm 1 i s }; {k\pm^{\prime \prime} 1  k 1 }  \right \}
\left \{ {j\pm^\prime 1 j t }; {k k\pm^{\prime  \prime}1 1 }\right \}
\nonumber \\
&&\times
{ \Theta(i,j,r) \Theta (j,k,t ) \Theta (i, k,s)  \over         
[r+1][s+1][t+1]   }
\end{eqnarray}

Here $\left \{ {i i i\pm1}; {112}  \right \}$ are the $6-j$ symbols,
and $\Theta(i,j,r) $ is the theta function defined in 
\cite{sn-lou,rsl,roumen-ham}.  The formula is written in
a way that is good for either the ordinary or $q$-deformed
case, so $[n]$ is the quantum integer \cite{sn-lou}, which is
equal to $n$ in the ordinary case.  

There is also the case in which there is in $\Gamma$ no
edge joining $v_1$ and $v_2$.  In this case the
operator adds one and gives it a color $1$.  
The topology of the edge is chosen so the loop
it forms with $e_1$ and $e_2$ links or intersects
no other edge of the network.
One then applies
the above formula  
with $\pm^{\prime\prime}=+$, and $k=0$, producing in this case
four terms.  

What this combinatorial formula corresponds to is adding a
loop as usual to represent the $F_{ab}$ in the plane of the
tangent vectors of the two edges.  The combinatorics are as in 
\cite{ham1,roumen-ham,ham2} except that the new loop
is taken to go around the triangle $e_1,e_2,e_{12}$.  

\item{\bf R3}  To complete the definition of the operator
we must divide by the area of the triangle $e_1,e_2,e_{12}$.
We may note that, as determined by the area operator, this will
often vanish, but it may instead be defined using 
Thiemann's length operator\cite{tt-length}  as follows.
If we call $\hat{L}_1,\hat{L}_2$ and $\hat{L}_{3}$ 
the length operators of
the edges of a triangle $\Delta$ of a spin-network, we may define an 
operator that measures its area as,
\f
\hat{A}_\Delta^2  = {1 \over 4}\left ( 
{ \hat{L}^2_1\hat{L}^2_2 \over 2}+{ \hat{L}^2_1\hat{L}^2_3 \over 2}
+{ \hat{L}^2_2\hat{L}^2_3 \over 2}
-{\hat{L}^4_1 \over 4}-{\hat{L}^4_2 \over 4}-{\hat{L}^4_3 \over 4}
\right )
\ff
 where we have used the standard formula from Euclidean geometry
for the area \footnote{We may note that postulating that 
a formula Euclidean
geometry holds in the microscopic level  is of course
justified only by the fact that it is the simplest unique choice.}.  (If the operators fail to commute we take symmetric
ordering.) 

We may note that this operator will generally 
yield a different answer than a 
direct measurement of the area, using the standard
area operator \cite{volume1}.  This 
is an inevitable consequence 
that we are working with a quantum field theory, in which
functional relationships between classical observables may not
be preserved.

We may then define this step as follows:  If there is a term
with no triangle corresponding to the three original edges we
do nothing.  If there is we multiply the state gotten by the first
two steps by the operator 
$\hat{A}^{-1}_{e_1,e_2,e_{12}}$\footnote{It is possible that 
there is a zero eigenvalue of the area.  To avoid this we need 
to define the
inverse so that those states do not contribute.  
We do so by defining $\hat{A}^{-1} \equiv \hat{A}^{-2} \hat{A}$,
where $\hat{A}^{-2}$ is defined on the subspace orthogonal
to the kernel of $\hat{A}$, 
so that terms that might contribute zero area are projected out.}

\item{\bf R4} Finally, we need to do something that corresponds to
integrating the constraint  against a lapse $N(x)$.  As we
are definining the operator combinatorically, we must make
an ansatz that is equivalent to doing this.  Using the criteria
that the action must agree on slowly varying-non-diffeomorphism
invariant states in the connection representation, we see that the
result of the usual definitions is to multiply by an indendent 
$N({v})$ at each node of a graph.  To define this combinatorically
we must make use of the recognition problem for subgraphs of
graphs.  We will multiply the action so far defined of the
operator on each node ${v}$ by numbers
$N({v})$, which are assumed to be assigned independently
to all nodes of all networks, subject to the following
restriction.  When it is the case that a network
$\Gamma$ may be identified as a subnetwork of $\Gamma^\prime$,
such that a given vertex $v$ of $\Gamma$ is identified uniquely
with a vertex $v^\prime$ of $\Gamma^\prime$
then $N(v)$ in the action on
$\Gamma$ must be taken equal to $N(v^\prime)$ of
$\Gamma^\prime$.

One might worry that there are many examples in which
a given graph $\Gamma$ may be identified in more than
one way with a subgraph of $\Gamma^\prime$, or
that the subgraphs may have symmetries that prevent the
unique identification of the node.  However, the combinatorics
of the graph recognition problem tells us that the proportion
of such cases goes to zero rapidly as the graphs become large.
As we are interested in the classical limit, and hence 
large complex graphs, this is sufficient.

\end{itemize}

The results of these four steps, applied to every node
of a basis element $\Gamma$ then gives a definition of
the operator ${\cal C}(N)$ on the space of diffeomorphism
invariant spin network states. 

It is easy to see that this definition eliminates the problem of
bounded correlations, so that a perturbation in a solution in
one region of a network will generally propagate over the whole.
We may note also that 
there are cases in which the adjacent nodes and
edges  are
eliminated by the action of the constraint.    For example
adjacent edges with color $1$ may be eliminated.  Also,
kinks in lines with color $1$ will be eliminated by the second
rule.   Thus, the repeated action of the operator
to any vertex will eventually produce terms that eliminate
either or both that vertex and its adjacent edges.  
As a result, the adjoint of the operator,
${\cal C}^\dagger$ will add nodes and edges.  

It is not known if the Hermitian form of this operator,
\ref{hermitian-form} has solutions, but if it does it
is then likely that they do not leave regions of a state
uncorrelated.  

Finally, we may note that there 
are still other approaches to the Hamiltonian
constraint in which there may emerge long
ranged correlations.  These include the approaches 
of \cite{BGP,GP,renata2}.
In fact, any approach that allows the Kodama
state\cite{kodama} 
(the exponential of the Chern-Simons invariant) as a
solution does generate long ranged correlations, as that state
is known to be both an exact physical state, with cosmological
constant, {\it and} a semiclassical state associated to DeSitter
spacetime\cite{chopinlee}.

\section{Concluding remarks}

In this paper we have described two problems that different 
formulations
of non-perturbative quantum gravity may suffer from.  We were 
able
to illustrate them with Theimann's formalism \cite{qsdi,qsdii} as it
allows a complete description of the space of solutions.  However
we saw also that at least one of the problems, that of the lack
of correlations which propagate over whole graphs, is likely
shared by other approaches.  Because of this we described a
new approach to the form of the Hamiltonian constraint, that
eliminates this difficulty, but at the cost of not following what
has become the canonnical procedure to construct diffeomorphism
invariant operators from point split regularization procedures.  

This is certainly progress.  But  we may wonder if it will be enough.
What if it is the case that a definition of the
Hamiltonian constraint that generates long-ranged correlations
is not enough to restore either a good classical limit or the
positivity of the $ADM$ energy? 

Indeed, while the existence of 
long-ranged correlations is a necessary condition for there
to be a description of the dynamics in terms of classical
geometry, there are reasons to think it may  not be sufficient.  
As mentioned in the introduction, 
experience with the renormalization group, random surface
theory and dynamical triangulations suggest that
to define a good continuum limit that reproduces classical
general relativity it is also necessary to tune the bare parameters
of the theory\cite{cosmocritic}.

We may note that in dynamical triangulations and Regge calculus the
 continuum limit only exists 
(if it exists at all) at a fixed point in the bare cosmological
constant-Newton's constant 
plane\cite{AM,Ambjorn,reggemodels,review-triangles}, 
where the bare
cosmological constant is nonvanishing.  
This suggests that at the very least, approaches to quantum
gravity that succeed without including a bare cosmological constant
may not have a good continuum limit, at least one that may be
related to any path integral description.  Of course, this follows
from general renormalization group considerations as well,
as one can generally never have a good continuum limit in a theory
without tuning the parameter of lowest dimension.

So the choice of a good hamiltonian constraint will most likely
need to be complemented by fine tuning of the bare Newton
and cosmological constants.  But 
what if even such fine tuning is not enough?   It may 
also be necessary
to introduce supersymmetry and other degrees of freedom,
in order to guarantee a good continuum limit.  Indeed,
it would not be surprising 
were this to turn out to be the case.  This might mean that in the end 
non-perturbative quantum gravity discovers conditions
for the existence of a continuum limit which are related
to those conditions known to be necessary  
for the existence of a sensible
perturbative quantum theory of gravity. (The point is that as far
as we know all such theories are string theories.)   
Supersymmetry will furthermore help with positive energy
as the $ADM$ operator becomes the square of the supersymmetry
generators.  It would be indeed interesting were fermionic
behavior, which is necessary to stabilize ordinary matter, 
also necessary to stabilize the quantum geometry of space.

Indeed, there
are only two alternatives to this scenario.  One is that 
non-perturbative quantum general relativity has by itself
a continuum limit whose perturbative description resembles
a supersymmetric string theory, which is the only
way we know to have a good perturbative description of
the interactions of gravitons.  The other is that
 a new perturbative description, so far
undiscovered, would have
to emerge from the continuum limit of non-perturbative
general relativity.

At the very least, the issues discussed here show that the problems of
the existence of the continuum limit and its
correspondence to the classical theory are key problems for
non-perturbative quantum gravity.    Further, these problems are
closely connected with the issues already discovered by other
discrete approaches such as dynamical triangulations and
Regge calculus\cite{AM,Ambjorn,reggemodels,review-triangles}.  
The moral of all of these stories is that a quantum
theory of gravity according to which Planck scale physics is
discrete can correspond to our world only if there is a natural
reason for the system to arrange itself into a critical state in which
correlation lengths diverge and massless particles may emerge.  

\section*{ACKNOWLEDGEMENTS}

I am grateful to Thomas Thiemann for correspondence about his
formulation of quantum gravity and for many helpful 
criticisms of the first draft of this paper.  Conversations and
correspondence with  Abhay Ashtekar, 
Stuart Kauffman, Jerzy Lewandowski, Fotini
Markopoulou, Don Marolf, Jorge Pullin, Michael Reisenberger
and Carlo Rovelli
were also most helpful. This work
has been supported by the NSF grant  PHY-93-96246.

\end{document}